\documentclass{llncs}
\usepackage[utf8]{inputenc}

\usepackage{url}
\usepackage{hyperref}

\usepackage{float}
\usepackage{caption} 
\usepackage{amsmath}
\usepackage{xcolor}
\usepackage{amsfonts}
\usepackage{verbatim}

\usepackage{csvsimple}
\usepackage{todonotes}

\hypersetup{colorlinks=true,linkcolor=blue, urlcolor=blue, citecolor=blue, linktocpage}
\newcommand{\doi}[1]{\url{http://doi.org/#1}}

\usepackage{graphicx}

\begin{document}
\title{New heuristic to choose a cylindrical algebraic decomposition variable ordering motivated by complexity analysis}
\author{Tereso del Río, Matthew England}
\institute{Coventry University, UK}
    
    \maketitle
    \email{delriot@uni.coventry.ac.uk, Matthew.England@coventry.ac.uk}
    
    \begin{abstract}

        It is well known that the variable ordering can be critical to the efficiency or even tractability of the cylindrical algebraic decomposition (CAD) algorithm. We propose new heuristics inspired by complexity analysis of CAD to choose the variable ordering. These heuristics are evaluated against existing heuristics with experiments on the SMT-LIB benchmarks using both existing performance metrics and a new metric we propose for the problem at hand. The best of these new heuristics chooses orderings that lead to timings on average 17\% slower than the virtual-best:  an improvement compared to the prior state-of-the-art which achieved timings 25\% slower.
      
    \end{abstract}

    \section{Introduction}\label{sec:introduction}

\subsection{Cylindrical algebraic decomposition}

    A Cylindrical Algebraic Decomposition (CAD) of $\mathbb{R}^n$ is a decomposition of $\mathbb{R}^n$ into semi-algebraic cells that are cylindrically arranged.  A cell being semi-algebraic means that it can be described by polynomial constraints.  CADs are defined relative to a variable ordering, for example, $x_n\succ x_{n-1}\succ \dots\succ {x_1}$.  Then the cylindrical property means that the projections of any two cells in $\mathbb{R}^n$ onto a subspace $\mathbb{R}^{i}, i<n$ with respect to this variable ordering, are either equal or disjoint.  I.e. the cells in $\mathbb{R}^n$ are arranged into cylinders above cells in $\mathbb{R}^{n-1}$, which are themselves arranged into cylinders above $\mathbb{R}^{n-2}$ and so on.  
    
    It can be very useful to find such decompositions satisfying a property such as sign-invariance for an input set of polynomials (i.e. each polynomial has constant sign in each cell).  The principle is that given an infinite space, a sign-invariant decomposition gives a finite set of regions on each of which our system of study has invariant behaviour, and thus can be analyzed by testing a single sample point.  When such a decomposition is also cylindrical and semi-algebraic we can use it to perform tasks like quantifier elimination.

    Collins in 1975 \cite{Collins1975} was the first to propose a feasible algorithm to build such sign-invariant decompositions for a given set of polynomials.  This algorithm has two phases, projection and lifting, each of them consisting of $n$ steps, where $n$ is the number of variables in the given set of polynomials $S_n$. 

    In the first step of the projection phase the given set of polynomials, $S_n$ is passed to a CAD projection operator to obtain a set of polynomials $S_{n-1}$ in $n-1$ variables (without the biggest variable $x_n$). This process is iterated until a set of polynomials $S_1$ only in the variable ${x_1}$ is left: at this point the projection phase ends.

    In the first step of the lifting phase a CAD of $\mathbb{R}^1$ is created by computing the $t$ ordered roots of $S_1$, denoted $r_1,\dots,r_t$, and building the CAD of $\mathbb{R}^1$ out of the cells $(-\infty,r_1), [r_1], (r_1,r_2),\dots, (r_{n-1},r_n),[r_n], (r_n,\infty)$. 
    Note that a sample point can be taken from each of those cells. 

    For each of those cells, the sample point is substituted into the set of polynomials $S_2$ to obtain a set of polynomials in one variable. Then using this set a stack of cells is built on top of each cell by following the instructions in the previous paragraph. These stacks are combined later into a CAD of $\mathbb{R}^2$ and by iterating this process a CAD of $\mathbb{R}^n$ is eventually built, concluding the lifting phase and the algorithm.
   
    The proof of correctness of CAD (allowing the conclusion of sign-invariance) relies on proving that the decompositions built over the sample point are representative of the behaviour over the entire cell:  to conclude this the projection operator must produce polynomials whose zeros indicate where the behaviour would change.  One representation of a CAD is as a tree of cells of increasing dimension; whose leaves are the cells in $\mathbb{R}^n$, nodes the cells in lower dimension, and branches representing the cylinders over projections.

    Since the introduction of CAD by Collins, many improvements have been made to the algorithm.  We do not detail them all here but refer the reader to the overview of the first 20 years in \cite{Collins1998} and to e.g. the introduction of \cite{Bradford2016} for some of the more recent advances.  We note in particular the recent developments in CAD projection \cite{McCallum2019} and the recent application of CAD technology within SMT and verification technology e.g. \cite{Kremer2020}, \cite{Abraham2021} which has inspired new adaptations of the CAD algorithm such as \cite{Brown2015}.  The work of this paper is presented for traditional CAD, but we expect it would transfer easily to these recent contexts.

\subsection{CAD variable ordering}

\begin{figure}[t]
	\centering
	\includegraphics[width=10cm]{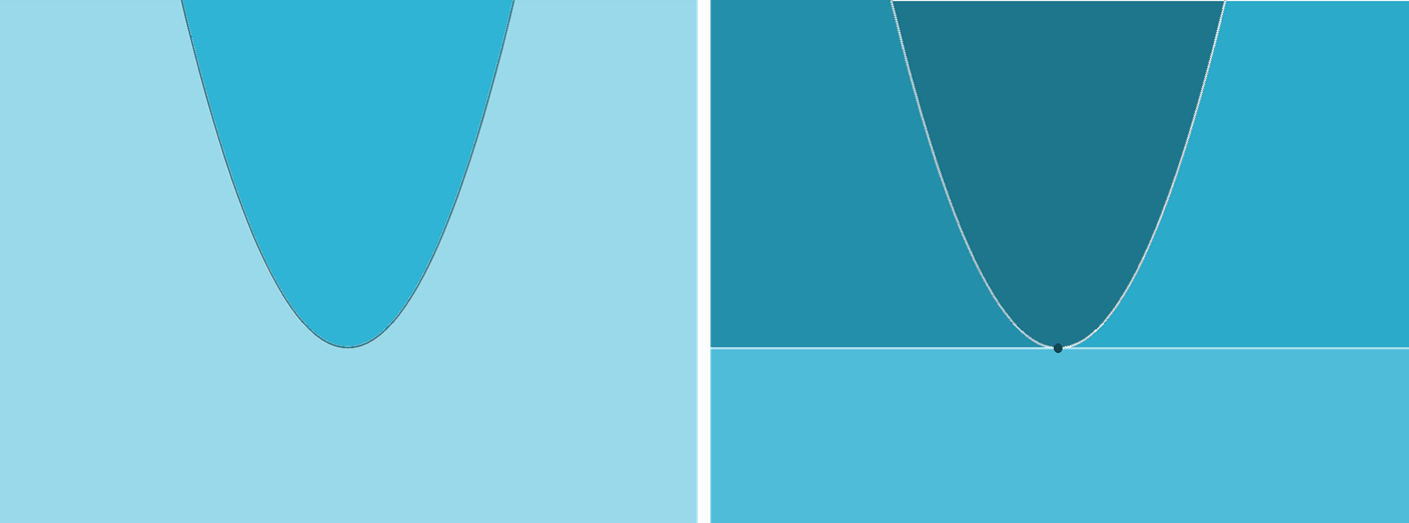}
		
	\caption{CADs sign-invariant for $x^2-y$ in the two possible orderings. The ordering $y\succ x$ generates the CAD on the left with only three cells (the coloured regions and the curve between).  The ordering $x\succ y$ generates the CAD on the right with nine cells (the four coloured regions along with four line segments and the turning point of the curve).  \label{fig:importanceordering}}
\end{figure}

It is well known that the variable ordering given can have a huge impact on the time and resources needed to build the CAD (see e.g. \cite{Dolzmann2004}, \cite{England2019}, \cite{Huang2019}).  We demonstrate this for a very simple example in Figure \ref{fig:importanceordering}, where one choice leads to three times the number of cells than the other.  In fact, \cite{Brown2007} shows that the choice of variable ordering can even change the theoretical complexity for certain classes of problems.

Depending on the application the CAD is to be used for, we have a free or constrained choice of variable ordering.  For example, to use a CAD for real quantifier elimination it is necessary to project variables in the order they are quantified, but there is freedom to swap the order of variables in quantifier blocks, and also to swap the order of the free (unquantified) variables and parameters.  Making use of this freedom is an important optimisation. This paper aims to present a new heuristic to pick the variable ordering for the construction of a CAD.

\subsection{Plan of the paper} 

    We continue in Section \ref{sec:previousheuristics} by describing the previous heuristics developed for the problem. Then in Section \ref{sec:proposedheuristics} the proposed heuristics are presented. We then move onto our experimental evaluation:  in Section \ref{sec:experimentsandbenchmarking} the methodology is detailed and in Section \ref{sec:analysis} the results obtained are analyzed. We finish with our conclusions and suggestions for future work in Section \ref{sec:final}.

    \section{Previous heuristics}\label{sec:previousheuristics}

    Due to how critical the variable ordering can be, a variety of heuristics have already been proposed for making the choice.  We will focus on two of these which are widely considered to constitute the current state-of-the-art:  the Brown heuristic, presented in \cite{Brown2004}, and \texttt{sotd}, presented in \cite{Dolzmann2004}.  We describe these in full over the coming subsections.  
    
    We acknowledge there are additional human-designed heuristics in the literature, but these are either even more expensive than \texttt{sotd} e.g. \cite{Bradford2013}, \cite{Wilson2015},
or designed relative to a very specific CAD implementation e.g.  \cite{England2014}.  

We also acknowledge that there exists a family of machine learning methods to take this decisions e.g. \cite{Huang2019}, \cite{England2019}, \cite{Florescu2019b}, \cite{Chen2020}, \cite{Brown2020} which have been shown to outperform the human-designed heuristics.  We do not compare against these directly but note that the lessons learnt in this paper could inform another generation of these machine learnt heuristics.

    \subsection{The Brown heuristic}\label{subsec:brown}


    The Brown heuristic was proposed by Brown in the notes to his ISSAC 2004 tutorial \cite[Section 5.2]{Brown2004}.  It chooses to project the variable with: 
    \begin{enumerate}
        \setlength\itemsep{0em}
        \item lowest degree; breaking ties with
        \item lowest value of the highest total degree term in which the variable appears; breaking ties with
        \item lowest number of terms containing the variable.
    \end{enumerate}
    Should there remain ties after the third measure then \cite{Brown2004} does not specify what to do: we name the variables $x_1,x_2,\dots$ according to the order in which they appear in the description of the problem and our implementation of the Brown heuristic chooses the variable with the lowest subindex first.
    
    The text in \cite{Brown2004} is also unclear on whether these measures are applied only to the input polynomials to produce the complete ordering, or whether they are applied to select an ordering one variable at a time, each time being applied to the polynomials obtained from projection with the last variable.  Our implementation does the latter:  this requires no additional projection computation above that required to build a single CAD, and matches the projection computation used by our new heuristic allowing for a fair comparison later.

For an example, let us consider the set 
$S_3=\{ {x_3}^3+{x_2}^3+{x_2}-{x_1}^4,{x_2}^3-{x_1}\}$.  
In $S_3$ the variable ${x_1}$ has degree $4$, but ${x_3}$ and ${x_2}$ both have degree $3$ so they tie in the first feature.  However, ${x_3}$ reaches this maximum in only one term, while ${x_2}$ does so in two of them and hence ${x_3}$ will be the first CAD projected variable. The CAD projection of $S_3$ with respect to ${x_3}$ is $S_2=\{ {x_2}^3+{x_2}-{x_1}^4,{x_2}^3-{x_1} \}$.  In $S_2$ the variable ${x_1}$ has degree $4$ while ${x_2}$ has degree $3$, so ${x_2}$ is the second variable to be projected.  This determines the variable ordering chosen with our implementation of the Brown heuristic: ${x_3}\succ {x_2}\succ {x_1}$.

    The motivation of the Brown heuristic is to try to make the next projected set of polynomials as small as possible.  The heuristic is ``cheap'' as the measures it uses require only easy to calculate characteristics of the input polynomials, and no algebraic computations such as projection.

    Moreover, the Brown heuristic has been shown to achieve similar accuracy to more complicated heuristics \cite{Huang2019}, \cite{Florescu2019a}, like the one introduced in the next section.  This means that when we include the cost of running the heuristics themselves, the Brown heuristic is actually superior.  

    Nevertheless, as will be seen later, it is possible to propose better heuristics by looking at the bigger picture rather than focusing on the next projected set.

    \subsection{The sotd heuristics}\label{subsec:sotd}

    The acronym \emph{sotd} stands for `sum of total degrees'. The heuristic \texttt{sotd} consists of computing the whole CAD projection of the input polynomials in every possible variable ordering and choosing the ordering whose projection has the smallest sum of total degrees throughout all the monomials in all the polynomials of the projection \cite{Dolzmann2004}. 

    For example, given the set of polynomials $S_3$ defined above and following variable ordering ${x_3}\succ {x_1}\succ {x_2}$, then we find $S_2$ as defined above and them projection with respect to ${x_1}$ gives $S_1=\{ {x_2}, {x_2}^2 + 1, {x_2}^{11} - {x_2}^2 - 1 \}$. The sum of all the degrees in the projected sets for this ordering is $43$.  It turns out that this is lowest such value possible from any of the six possible orderings.  Hence the \texttt{sotd} heuristic would choose this variable ordering.
    
    This heuristic is not very desirable at first glance, because $\mathcal{O}(n!)$ 
    projection steps are needed to make the choice, far more than the $n$ projections used to make a single CAD.    
    This problem was spotted by the authors of \cite{Dolzmann2004} leading them to develop a ``greedy'' version of the heuristic in the same paper. This was greedy in the sense that instead of selecting the entire ordering at once using information from the whole projection phase for each possible ordering; it computed the ordering one variable at a time by comparing their metric on the output of a single projection step for each possible variable from the remaining ones. This version only requires $\frac{n(n+1)}{2}$ number of projections, still more than the amount of projection normally used to build a CAD.  In our experiments, the choices it makes are much poorer than those made by \texttt{sotd} with the full projection information.  In fact, some experiments show it even performs worse than the Brown heuristic, despite having more projection information available to make its choice.  We see this later in our experiments (Table \ref{tab:allvsallInf0}).

    The metric \textit{sotd}, i.e. summing all the degrees in a set of polynomials, was originally constructed as a measure of the overall size of a polynomial set.  In \cite{Dolzmann2004} it was used along with other measures to demonstrate the effect of variable ordering on CAD computation.  It was found to have a strong correlation with those other CAD complexity measures, but unlike them, it did not require the computation of the entire CAD.  This led to its proposal for using it as a heuristic.   
    
    Thus, it seems the \texttt{sotd} measure was not designed primarily for use in a CAD variable ordering heuristics,  allowing a gap for the new results of the present paper which presents a measure that is designed this way.  This lack of tailoring to CAD can be seen in the way \texttt{sotd} and its greedy version give the same importance to all degrees found in the projected sets of polynomials regardless of the variable carrying that degree, whereas CAD works iteratively one variable at a time (treating the others as part of the coefficients when projection and substituting them for the sample point when lifting).  Thus different variables carry different weights of effect on CAD computation. 

    In the next section, it will be shown how better heuristics can be proposed which take into account the potential growth in complexity of the polynomials we observe in CAD projection.

    \section{Our new proposed heuristics}\label{sec:proposedheuristics}

    It has been shown in \cite{Dolzmann2004} that the number of cells of a CAD is strongly correlated with the time taken to build that CAD.  
    Hence the most recent CAD complexity analyses in e.g. \cite{England2015}, \cite{Bradford2016}, \cite{England2020}, \cite{Li2021} have studied a bound on the maximum number of cells that can be generated.

    The idea of the proposed heuristic is to make a corresponding estimate on this maximum number of cells of the final CAD for each of the possible orderings and pick the ordering that minimizes this value. 
    
    We explain how this number can be computed or estimated if the whole CAD projection is known for each ordering. However, as CAD projection of polynomials can be an expensive operation, a greedy version of this heuristic that does not require any CAD projections to make the choice is also proposed.

    \subsection{Heuristic motivated by a complexity analysis: \texttt{mods}}\label{subsec:heuristicbycomplexity}

    Define the \emph{degree sum} of a variable $x$ in a set of polynomials $S=\{p_1,\dots,p_n\}$ as 
    \begin{equation}\label{eq:degreesum}
        D_x(S)=\sum_{i=1}^n d_x(p_i),
    \end{equation}
    where $d_x(p)$ is the degree of $x$ in the polynomial $p$.  Thus the maximum number of unique real roots that the polynomials in $S$ can have with respect to $x$, is $D_x(S)$.

    To compute a CAD with the variable ordering $x_{n}\succ\dots\succ x_{1}$ for the set of polynomials $S_n$, the set $S_{n}$ must be projected with respect to $x_{n}$ to obtain the set $S_{n-1}$; and in the same fashion the sets $S_{n-2},S_{n-3},\dots,S_1$ are computed.
    
    Thus when following the creation of a CAD as described in Section \ref{sec:introduction}, in the first lifting step at most $2D_{{x_1}}(S_1)+1$ cells can be created because ${x_1}$ will have at most $D_{{x_1}}(S_1)$ roots in $S_1$. Subsequently at most $(2D_{{x_1}}(S_1)+1)\cdot (2D_{x_2}(S_2)+1)$ cells will be built in the second lifting step (a similar limit applied for each stack above a cell from $\mathbb{R}^1$).  
    
    Hence, at the end of the lifting phase, when the CAD is completed, an upper bound on the number of cells is 
    \begin{equation}\label{eq:maxcellsinfinalCAD}
        \prod_{i=1}^{n} \big( 2D_{x_i}(S_i)+1 \big).
    \end{equation}

    As discussed earlier, the number of cells in a CAD is strongly correlated with the time needed to build such CAD. Therefore, choosing the ordering that minimizes the maximum number of cells in the final CAD sounds like a good idea if we want to choose a fast ordering. Hence, we want to choose the ordering that minimizes \eqref{eq:maxcellsinfinalCAD}.

    We note that the dominant term of \eqref{eq:maxcellsinfinalCAD} is  
        \begin{equation}\label{eq:mods}
            \prod_{i=1}^{n} D_{x_i}(S_i),
        \end{equation} 
    which we refer to as the \emph{\textbf{m}ultiplication \textbf{o}f \textbf{d}egree \textbf{s}um (mods)}.  By minimising \eqref{eq:maxcellsinfinalCAD} we are likely minimising this and so  we refer to the heuristic that picks an ordering to minimise \eqref{eq:maxcellsinfinalCAD} as \texttt{mods}. As with our implementation of the Brown heuristic, we apply this to choose one variable at a time, projecting with respect to that variable after the choice and then applying the measure to the projection polynomials to make the next choice.  In case where there is a tie on the measure then we pick the variable with the lowest subindex.

    Consider our example set of polynomials $S_3=\{{x_3}^3+{x_2}^3+{x_2}-{x_1}^4,{x_2}^3-{x_1}\}$. The degree sum of ${x_1}$ in $S_3$ is 
\[
    D_{x_e}(S_3)=d_{x_e}({x_3}^3+{x_2}^3+{x_2}-{x_1}^4) + d_{x_e}({x_2}^3-{x_1}) = 3+0 = 3.
\]    
Suppose we built a CAD using the variable ordering ${x_3}\succ {x_1}\succ {x_2}$.  Then as before we obtain $S_2=\{-{x_2}^3 + {x_1}, {x_1}^4 - {x_2}^3 - {x_2}\}$ for which $D_{{x_1}}(S_2)=5$, and $S_1=\{{x_2},{x_2}^2+1,{x_2}^{11}-{x_2}^2-1\}$ for which $D_{x_2}(S_1)=14$. Therefore, for this example CAD the product \eqref{eq:maxcellsinfinalCAD} evaluates to 2233.  It turns out that is the lowest value of \eqref{eq:maxcellsinfinalCAD} for all the possible orderings.  Hence \texttt{mods} would have chosen this ordering.

    \subsection{Creating a greedy version of \texttt{mods}}\label{subsec:proposedgreedyheuristic}

    As with \texttt{sotd}, the heuristic \texttt{mods} is relatively expensive, requiring the use of CAD projection operations in all different variable orderings.  To reduce its cost we present a greedy version of this heuristic, that will simply choose to project the variable with the lowest degree sum (see (\eqref{eq:degreesum})) in the set of polynomials\footnote{We note that this measure applied only to the original polynomials is one of the features that was generated algorithmically to train different machine learning classifiers to take decisions on sets of polynomials in \cite{Florescu2019a}.}
    
    This heuristic will be referred to as \texttt{gmods}. Note that unlike \texttt{greedy-sotd}, \texttt{gmods} does not use any projection information beyond that required to build a single CAD.  The metric it is based on uses only easily extracted information from the polynomials.  It is thus similar in cost to our implementation of the Brown heuristic.

    For example, given the set of polynomials $S_3$ above we have $D_{x_1}(S_3)=5$, $D_{x_2}(S_3)=6$ and $D_{x_3}(S_3)=3$.  Thus \texttt{gmods} will select ${x_3}$ as the first variable for CAD projection.  The CAD projection of $S_3$ with respect to ${x_3}$ gives $S_2$ as above.  In $S_2$ the variable ${x_1}$ has degree sum $5$ while ${x_2}$ has degree sum $6$, so ${x_1}$ is the second variable to be projected, determining completely the variable ordering that \texttt{gmods} chooses: ${x_3}\succ {x_1}\succ {x_2}$.

    \subsection{Heuristic motivated by expected number of cells}

    Our \texttt{mods} heuristic is motivated to reduce the maximum number of cells that could be computed according to a complexity analysis.  It is natural to ask whether we could be more accurate and seek take decisions according to an expected value of the number of cells rather than the maximum?
    
    To calculate the maximum number of cells that can be generated, the degree of the polynomials has been used because it is the maximum number of real roots that a polynomial can have, so we may consider the expected number of roots of a polynomial.  According to \cite{Fairley1976},
    the expected number of real roots for polynomials of small degree is proportional to the logarithm of its degree, at least for their definition of random polynomials.  However, for a linear polynomial, this relation would predict zero roots when it should be one, and so to address this we suggest a heuristic following this approach should add one before taking the logarithm.  
    
    Thus we hypothesise an expected number of cells in the final CAD as below (following the approach of Section \ref{subsec:heuristicbycomplexity}): 
    \begin{equation}\label{eq:logmods}
        \prod_{i=1}^{n} (2\log(D_{x_i}(S_i)+1) + 1). 
    \end{equation}
   As before, we define a heuristic to pick the ordering that minimizes \eqref{eq:logmods}.  Given the similarity to \texttt{mods} and the use of the logarithm we refer to this as \texttt{logmods}.

   For example, consider the set of polynomials $S_3$ as before and the variable ordering ${x_3}\succ {x_1}\succ {x_2}$ to produce $S_2$ and $S_1$ as before.  We find $D_{x_3}(S_3)=3$, $D_{x_1}(S_2)=5$, and $D_{x_2}(S_1)=14$. Therefore, \eqref{eq:logmods} evaluates to $15.43$, and it turns out that is the lowest value for all possible orderings, hence, \texttt{logmods} would choose this variable ordering.

    \section{Experiments and benchmarking}\label{sec:experimentsandbenchmarking}

    \subsection{Benchmarking}

    The three-variable problems in the QF\_NRA category of the SMT-LIB \cite{Barrett2016} are used to build a dataset for comparing the different heuristics. 

    For each of those problems and all possible orderings, we timed (see Section \ref{subsubsec:timing}) how long it takes to build a sign-invariant CAD for the polynomials involved, discarding the problems in which the creation of the CAD timed out for all orderings.
    After building all the possible CADs, a dataset of ``unique'' problems is created (see Section  \ref{subsubsec:uniqueness}).

    Of the 5942 original problems, in 343 of them, all the orderings timed out. And out of the remaining 5599 problems, only 1019 unique problems were found. These 1019 problems will be used as benchmarks to compare the heuristics presented in Sections \ref{sec:previousheuristics} and \ref{sec:proposedheuristics}.

	\subsubsection{CAD Implementation}
	
	For our experiments we used the function \\ \texttt{CylindricalAlgebraicDecompose} in the \textsc{Maple} 2022 Library \texttt{RegularChains}, whose implementation is  described in \cite{Chen2014}.  This actually implements a somewhat different CAD algorithm to the classical approach described above.  Instead of projecting and lifting it first decomposes complex space and then refines this to a CAD \cite{Chen2009}, with the current implementation doing the complex decomposition incrementally by polynomial \cite{Chen2014a}. As reported in these papers, this approach can avoid some superfluous cell divisions.  However, there is still the same choice of variable ordering to be made which can be crucial \cite{Chen2020} with the Brown heuristic observed previously to work similarly well for the regular chains based algorithms \cite{Huang2019}.  
	
    \subsubsection{Timings}\label{subsubsec:timing}
    
    Timings are performed following the methodology of \cite{England2019}. For each of the possible variable orderings, the polynomials defining the problems were given as input to the CAD in \textsc{Maple} with a time limit of 30 seconds. If none of the orderings finishes, all the orderings are attempted again with a time limit of 60 seconds.

    Projection times are timed individually using our implementation in \textsc{Maple} of McCallum CAD projection (that returns the polynomials factorized) with a time limit of 10 seconds: these times are used to give a more meaningful comparison of heuristics that requires us to compute all the projections, with heuristics that do not need to do so.

    Every CAD call was made in a separate \textsc{Maple} session launched from and timed in Python, to avoid \textsc{Maple}'s caching of intermediate results from one benchmark or ordering that may help another. From each timing, 0.075 seconds were removed: the average time that \textsc{Maple} takes to open on the computer when called from Python.  It was removed as this is not a cost that would normally be paid but as a consequence of the benchmarking.

    \subsubsection{Uniqueness}\label{subsubsec:uniqueness}
    
    When studying the dataset it was observed that many examples were very similar to each other. Similar in the sense that they were described by very similar polynomials, resulting in CADs with equivalent tree structures for every variable ordering, making it likely that all aspects of the CAD generation were similar.  It is well observed that there exist these families of very similar benchmarks in the SMT-LIB.  Treating each of them as an independent benchmark could result in skewed experimental results. E.g. a heuristic that happens to perform well on a large family of almost identical benchmarks would receive a huge but unwarranted boost in the analysis if we do not take care.  
    
    To avoid this, the samples with the same number of cells in the CADs for all possible variable orderings are clustered and only one of them is included in the dataset. This ensures that there are no two problems with an equivalent CAD tree structure for each variable ordering.

    \subsection{Evaluation metrics}\label{subsec:metrics}

\subsubsection{Existing evaluation metrics}

    The most obvious metric to evaluate the choices of our heuristics is the total time taken to build CADs for all the problems with the orderings chosen by that heuristic: this metric will be referred to as total-time. Also, another metric that will be used to compare the different heuristics is the number of problems completed before timeout using the orderings chosen by the heuristic.
    
    In previous studies such as \cite{Florescu2019a} accuracy, i.e. the percentage of times that the fastest ordering is chosen, is used as one of the main metrics.  However, as discussed in \cite{Florescu2019b}, for our context, accuracy is not the most meaningful metric.  This is because it is well observed that the second-best ordering may only be very marginally worse than the best ordering and so picking that should also be considered accurate.  Further, the timings may include small amounts of computational noise which change the ranking of orderings in such subsets and thus the accuracy score.  
    
    In \cite{Florescu2019b} the authors proposed to address this by considering a heuristic as successful if it identifies any ordering that takes no more than 20\% additional time than the optimal.  This fitted their work on a machine learning classification problem, but this definition is not suitable for regression, or use to evaluate a continuous range of possibilities.  It considers equally inaccurate an ordering that is 30\% slower and an ordering that is three or four times slower, and even an ordering that timed out.  We thus propose a new metric for use in the evaluation in place of accuracy.
    
\subsubsection{Markup}
    
    We suggest measuring the amount of time that the chosen ordering takes above the time of the optimal ordering, as a percentage of the optimal ordering:
    $$
    \frac{ \rm{heuristic}\_\rm{time} - \rm{optimal}\_\rm{time} }        
         { \rm{optimal}\_\rm{time}}.
    $$
    This allows for problems of different sizes to be evaluated relative to their possible solutions.  For example, suppose Problem A's optimal ordering took 10 seconds and Problem B's took 20s.  If the chosen ordering for Problem A took 2s longer than optimal then the score would be 0.2; while if that happened for Problem B the score would be 0.1, recognizing that the excess 2s is a less substantial markup for the larger problem.
    
    However, this can lead to distortions for problems where the optimal ordering is really fast.  For example, if the optimal ordering takes 0.02s and the chosen ordering takes 4s then the metric above would give that problem a very huge influence over the final score. To avoid that situation, and taking into consideration that anything below a second would likely be acceptable to use for constructing a CAD, we propose instead to add one to all the timings, i.e.
    $$
    \rm{Markup} = \frac{(\rm{heuristic}\_\rm{time} + 1) -(\rm{optimal}\_\rm{time}+1)}{\rm{optimal}\_\rm{time}+1}.
    $$
This measure still allows the evaluation of relative potential but reduces distortions from fast examples and computational noise.  In the example above, the metric would evaluate to 3.9 instead of 199.  We refer to this as \emph{Markup}, i.e. a measure of how far from the optimal this choice was.

Markup combines the benefits of both accuracy and total-time. Like total-time does it can measure not only if a choice was worse than the virtual-best but also how worse it was.  But it adapts better to the different sizes of examples, unlike total-time where performing slightly worse in a difficult problem can have more impact on the metric than performing really bad in an easy example. Like accuracy it gives the same relevance to all the instances, but unlike accuracy it does not define a choice as simply either right or wrong.



 
\subsubsection{Timeouts} 
  
For computing 
markup and total-time we must decide how to deal with cases where the chosen variable ordering leads to a timeout in CAD computation.  In this case, when an ordering does not finish within the time limit given it will be assumed that it would have taken twice the time limit given
.

\subsection{Metrics and expensive heuristics}

Note that some of our heuristics are cheap, manipulating over data easily extracted from the polynomial, while others are expensive, requiring the use of CAD projection and thus algebraic computations.  When analyzing an expensive heuristic we have the choice of ignoring the cost of the heuristic or taking it into account. It is clear that the latter is more realistic because without paying this cost it would not be possible to make the choice. However, the former way of analyzing the heuristic also brings some interesting insight. Therefore, when presenting the metrics for these heuristics (Table \ref{tab:allvsallInf0}), the metric without including the cost of the heuristic will be shown between brackets.

For example, the number of examples marked as complete stands for the number of problems in which the CAD was constructed with the heuristic's choice of variable ordering before the timeout.  To adjust this in expensive heuristics, we count as timeouts the problems in which the time taken to choose the ordering plus the time taken to build the CAD did not exceed the time limit.  As the more realistic value, the latter is outside brackets and the former within.

    \section{Results and analysis} \label{sec:analysis}

\begin{figure}[H]
            \centering
            \includegraphics[width=10cm]{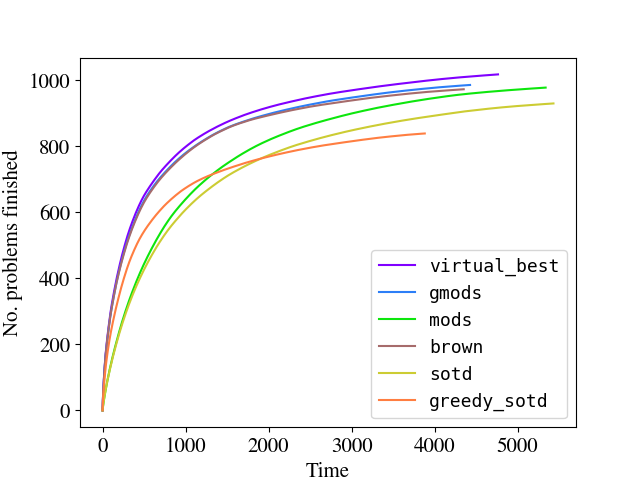}           
            \caption{Survival plot comparing heuristics on the benchmarks they can tackle before timeout. \label{fig:survivalmin0} }
        \end{figure}

    The results given by the analysis of the different heuristics to choose the variable ordering for the 1019 benchmarks are summarized in Table \ref{tab:allvsallInf0}, and a survival plot comparing the heuristics is presented in Figure \ref{fig:survivalmin0}.  
    To produce the survival plot, for each heuristic the times taken to solve the problems with the variable ordering chosen by the heuristic are sorted into increasing order to form a sequence $(t_i)$, discarding the timed-out problems; and the points $(k, \sum_{i=1}^k t_i)$ are then plotted. This plot encapsulates visually a lot of information about the success of the heuristics on a given dataset (it does not say anything about heuristics relative performance on particular problem instances).

    \begin{table}[H]
        \def\arraystretch{1.5}
        \setlength\tabcolsep{0.2cm}
        \begin{tabular}{|l|l|l|l|l|l|}\hline%
            \bfseries Name & \bfseries Accuracy & \bfseries Total time & \bfseries Markup
             & \bfseries \# Completed
            \csvreader[
            head to column names,
            filter=\equal{\Name}{sotd} \or \equal{\Name}{mods} \or \equal{\Name}{gmods} \or \equal{\Name}{brown} \or \equal{\Name}{random} \or \equal{\Name}{virtual-best} \or \equal{\Name}{greedy-sotd} \or \equal{\Name}{logmods}
            ]{
            study_heuristics_guess__without_repetition__max_penalisation_Inf__min_time_0.csv}{}{%
            \ifcsvstrequal{\Name}{sotd}
            {\\\hline \texttt{\Name} & \Accuracy & \TotalTime (\TotalTimeWithoutCost) & \Markup (\MarkupWithoutCost) 
            & \Terminating (\TerminatingWithoutCost)}
            {
            \ifcsvstrequal{\Name}{mods}
            {\\\hline \texttt{\Name} & \textbf{\Accuracy} & \TotalTime (\textbf{\TotalTimeWithoutCost}) & \Markup (\textbf{\MarkupWithoutCost})
             & \Terminating (\textbf{\TerminatingWithoutCost})}
            {
            \ifcsvstrequal{\Name}{greedy-sotd}
            {\\\hline \texttt{\Name} & \Accuracy & \TotalTime (\TotalTimeWithoutCost) & \Markup (\MarkupWithoutCost) 
            & \Terminating (\TerminatingWithoutCost)}
            { 
            \ifcsvstrequal{\Name}{logmods}
            {\\\hline \texttt{\Name} & \Accuracy & \TotalTime (\TotalTimeWithoutCost) & \Markup (\MarkupWithoutCost) 
            & \Terminating (\TerminatingWithoutCost)}
            {
            \ifcsvstrequal{\Name}{gmods}
            {\\\hline \texttt{\Name} & \Accuracy & \textbf{\TotalTime} & \textbf{\Markup} 
            & \textbf{\Terminating}}
            {\\\hline \texttt{\Name} & \Accuracy & \TotalTime & \Markup 
            & \Terminating}
            }
            }
            }
            }
            }%
            \\\hline
        \end{tabular}

        \caption{Evaluation metrics for the different heuristics to choose the variable orderings for CAD. For the expensive heuristics, the metrics without taking into account the cost of the heuristic can be seen between brackets. In bold, the best measure of the metric out of all the heuristics. \label{tab:allvsallInf0} }
        \end{table}

The first thing to note is that \emph{any} heuristic is significantly better than making a random choice, giving further evidence on the critical need for attention to this decision.  We give further analysis on the heuristics grouped by their relative costs.

\subsection{Expensive heuristics: \texttt{sotd} vs \texttt{mods}}\label{subsec:expensiveheuristics}

    As discussed earlier, one of the strategies that can be taken to choose an ordering is to compute all the projection phases for each variable ordering beforehand and base the decision on this information. Our new heuristic \texttt{mods} and the existing heuristic \texttt{sotd} follow this strategy.  This approach requires a huge cost:  it is possible to observe in Table \ref{tab:allvsallInf0} that one-fifth and one-seventh of the time taken to do a CAD by the ordering suggested by \texttt{mods} and \texttt{sotd} respectively is invested on deciding the ordering.
    
    They both use the same information to take the decision, however, as can be seen in Table \ref{tab:allvsallInf0}, \texttt{mods} outperforms \texttt{sotd} in all the presented metrics. 
    The proposed heuristic picks the best ordering for almost two-thirds of the problems, while the existing one does so in less than half of them. The choices of \texttt{mods} reduce the total time by almost fifty minutes and solve almost 50 problems more with respect to the choices done by \texttt{sotd}. 

    Moreover, the ordering chosen by \texttt{mods} took on average 58\% more time than the best ordering, while the choice of \texttt{sotd} took on average 160\% more than it. Meaning that for a problem where the virtual-best ordering takes 10 seconds it is expected that the ordering proposed by  \texttt{mods} takes 15.8 seconds while 26 seconds are expected from the ordering proposed by \texttt{sotd}.

    When we compare \texttt{mods} to \texttt{logmods} we see that \texttt{logmods} is outperformed in all measured metrics.  We thus conclude that the expected number of real roots for the polynomials in our benchmark set does not match well that of the random polynomials studied in \cite{Fairley1976}.

\subsection{Cheaper heuristics:  \texttt{gmods} vs \texttt{brown}}

The heuristics presented in Section \ref{subsec:expensiveheuristics} exerted a large amount of effort to solve the problem of choosing the ordering, at odds with the behaviour of most algorithm optimization heuristics. 

We now look at the cheaper heuristics:  \texttt{greedy-sotd} which greatly reduces the amount of projection information used to make a decision (compared to \texttt{sotd} and \texttt{mods}) and \texttt{brown} and \texttt{gmods} which do not use any such information beyond that required to build the single CAD.

We first note that even without taking into account the higher cost of\\  \texttt{greedy-sotd}, it is greatly outperformed by the two heuristics that do not make use of projection information at all.  When comparing \texttt{brown} and \texttt{gmods}:  both heuristics have similar accuracies, however, the choices of \texttt{gmods} reduce the total time by ten minutes, and solves 13 problems more with respect to the choices of \texttt{brown}.  Moreover, the ordering chosen by \texttt{gmods} took on average 17\% more time than the best ordering, while the choice of \texttt{brown} took on average 25\% more than it. Meaning that for a problem where the optimal ordering takes 10 seconds it is expected that the ordering proposed by  \texttt{gmods} takes 11.7 seconds while 12.5 seconds are expected from the ordering proposed by \texttt{brown}. 

\begin{figure}[H]
        \centering
        \includegraphics[width=10cm]{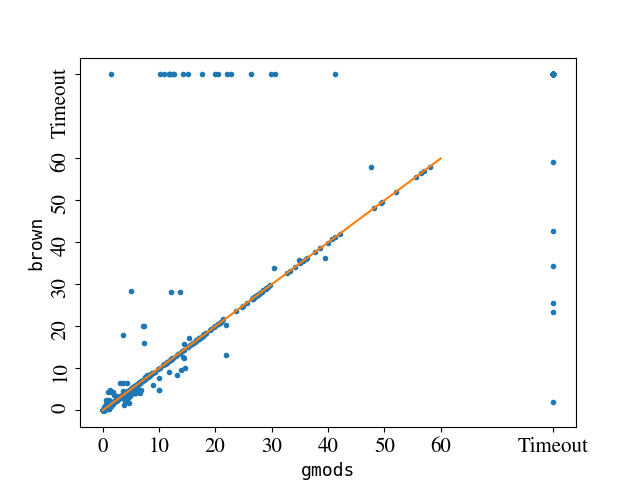}
        
        \caption{Adversarial plot comparing \texttt{gmods} and \texttt{brown}.  \label{fig:gmodsvsbrown}}
    \end{figure}

To further understand how these two heuristics compare, and if there are subsets of problems in which one of them performs better than the other, an adversarial plot is presented in Figure \ref{fig:gmodsvsbrown}. This plots for each benchmark the time taken by the two heuristics against each other.  In that figure, it can be observed that most of the points are close to the diagonal line, implying that both heuristics perform similarly for most of the instances.

However, it can be also observed that for some problems the ordering suggested by \texttt{gmods} timed out while \texttt{brown} proposed an ordering that completes and vice versa.  This phenomenon is more common in favour of \texttt{gmods} but leaves open the possibility of a combination or meta-heuristic outperforming either.

\subsection{Expensive vs cheap approach: \texttt{mods} vs \texttt{gmods}}

    It can be observed looking at the accuracies in Table \ref{tab:allvsallInf0} that \texttt{mods} picks better orderings than \texttt{gmods} and is superior in all other metrics if the cost of the heuristic is not taken into account (looking at the values between brackets in the same table).  I.e. it has the strongest predictive power.  However, as discussed in Section \ref{subsec:metrics}, accuracy is not the most interesting metric and \texttt{gmods} outperforms \texttt{mods} in all fair comparisons that take the cost of the heuristic into account.

    In Figure \ref{fig:survivalmin0} we see that \texttt{greedy-sotd} outperforms \texttt{sotd} at first, solving many problems in a shorter time, but in the long run \texttt{sotd} ends up solving more in total. In fact, it is possible to observe that the greedy heuristics start ahead of the expensive heuristics.  This implies that it is especially disadvantageous to compute all projections when working on easy problems, and motivates a separate analysis excluding the easiest problems.

\begin{figure}[H]
        \centering
        \includegraphics[width=10cm]{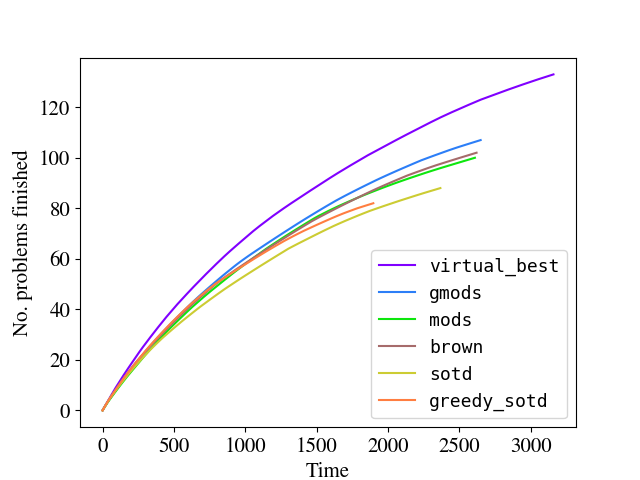}
        
        \caption{Survival plot comparing heuristics on harder benchmarks only.   \label{fig:survivalmin10}}
    \end{figure}

    The results when we restrict to only the hardest 134 problems (those whose optimal time need more than $10$ seconds) are plotted in Figure \ref{fig:survivalmin10}. Now \texttt{mods} performs almost as well as \texttt{brown}. This further highlights the superiority of \texttt{gmods} over the rest in this particularly relevant slice of the problems where easy problems are excluded. Thus these expensive heuristics may still have a role as we expand our analysis to still harder problems.

\section{Final thoughts}
\label{sec:final}

    \subsection{Conclusions}\label{sec:conclusions}

    The new heuristics motivated in this paper by the complexity analysis of CAD have clearly become the new state-of-the-art to choose the variable ordering for CAD. This leads to the most important conclusion of the paper: theoretical complexity analyses of an algorithm are a very powerful tool not just to compare algorithms but also to optimize them. These results also show that the benefits of a greedy or lazy approach in algebraic computation.  
    
    We would also highlight how out of the 5599 problems of three variables in the QF\_NRA category of the SMTLIB library, only 1019 were found to be unique (in the sense of \ref{subsubsec:uniqueness}).  I.e. three-quarters, at least of the three-variable problems in the QF\_NRA category, are from the point of view of CAD copies of other problems in that category.  Their differences may be important when using SAT solvers to study the logic, or other incomplete methods, but for an analysis of a complete solver they should be merged as in our methodology.

    Finally, we note that \texttt{logmods} failed in our experiments, implying that for the polynomials found in the QF\_NRA category of the SMTLIB library the expected number of roots is not proportional to the logarithm of their degree, and therefore they do not follow the same distribution as studied in \cite{Fairley1976}.  This is not that surprising given it is well documented that polynomials from applications are often different to those generated from a simple random generator.

    \subsection{Future work}\label{sec:futurework}

    An obvious future work is to experiment with the new heuristics on problems with more variables, higher degree, or data from other sources.  We see a number of avenues beyond this for more research.

    First we note that the results here should feed into work on machine learning methods to make such optimization decision.  In fact, it would be interesting to study the extend to which the metrics presented here were used in the machine learning classifiers of \cite{Florescu2019a} who created similar metrics among hundreds via an automated method.  
    We are also interested in building simpler ML models using a restricted set of features.  These could be more interpretable and thus offer further insights on how the features connect.

    Next, we note that the actual CAD complexity analyses in e.g. \cite{England2015}, \cite{Bradford2016}, \cite{England2020} \cite{Li2021} were performed not on the projection polynomials as a single set but an optimal arrangement of them.  This is known as the (m,d)-property and stems from the PhD thesis of McCallum.  It would be interesting to see if there is any heuristic that can be deduced from the analysis involving this property.

    We are also interested to look if the additional information encoded in \texttt{mods} could be obtained more cheaply than using CAD projections.  Especially since the heuristic does not use all the information in these projections, only degree information. This would allow us to make choices only 13\% slower on average than the virtual best (based on the results between brackets $-$ without heuristic cost $-$ in Table \ref{tab:allvsallInf0}).  Alternatively, another greedy version of \texttt{mods} could be developed, in which, similarly to \texttt{greedy-sotd}, instead of computing the whole projection phase for each possible variable ordering, a projection step is done for each available variable.

\section*{Acknowledgements}

The authors would like to thank AmirHosein Sadeghimanesh for his interesting conversations and his constructive criticism, and for sharing his Maple code to perform CAD projections.  We also thank the anonymous reviewers whose comments helped us improve the paper.

The research of the first author is supported financially from a scholarship of Coventry University.  The research of the second author is supported by EPSRC Grant EP/T015748/1, \emph{Pushing Back the Doubly-Exponential Wall of Cylindrical Algebraic Decomposition} (the DEWCAD Project).

\section*{Research Data and Code Statement}

Data and code necessary to generate the figures and results presented in this paper are available at: \url{https://doi.org/10.5281/zenodo.6750528}

    \bibliographystyle{../../styles_and_libraries/splncs04}

\end{document}